\documentclass[aps,amsmath,amssymb, reprint, prl, preprintnumbers, twocolumn]{revtex4-2}
\pdfoutput=1

\usepackage{graphicx}
\usepackage{epstopdf}
\usepackage{amsmath, amssymb}

\usepackage{hyperref}
\usepackage{bbm,array,amsfonts,graphicx,wrapfig,float,mathtools,multirow}
\usepackage[dvipsnames]{xcolor}

\usepackage{tikz, float}
\usetikzlibrary{patterns,shapes.misc}

\newcommand{\be}{\begin{equation}}
\newcommand{\ee}{\end{equation}}
\newcommand{\beq}{\begin{equation}}
\newcommand{\beql}[1]{\begin{equation}\label{#1}}
\newcommand{\eeq}{\end{equation}}
\newcommand{\ba}{\begin{array}}
\newcommand{\ea}{\end{array}}
\newcommand{\bea}{\begin{eqnarray}}
\newcommand{\beal}[1]{\begin{eqnarray}\label{#1}}
\newcommand{\eea}{\end{eqnarray}}
\newcommand{\ben}{\begin{enumerate}}
\newcommand{\een}{\end{enumerate}}
\newcommand{\bean}{\begin{eqnarray*}}
\newcommand{\eean}{\end{eqnarray*}}
\newcommand{\eref}[1]{(\ref{#1})}

\newcommand{\tref}[1]{Table~\ref{#1}}
\newcommand{\nn}{\nonumber}

\newcommand{\fref}[1]{Figure \ref{#1}}

\begin{document}

\title{Diagnosing Inconsistencies in $4d$ $\mathcal{N}=1$ Gauge Theories with Explainable AI}

\preprint{UNIST-MTH-26-RS-06}
\preprint{CGP26020}

\author{Seong-Jin Lee}
\email[\texttt{seongjinlee@ibs.re.kr}]{}
\affiliation{
Center for Geometry and Physics, Institute for Basic Science (IBS),
Pohang 37673, South Korea
}

\author{Rak-Kyeong Seong}
\email[\texttt{seong@unist.ac.kr}]{}
\affiliation{
Department of Mathematical Sciences, and Department of Physics, Ulsan National Institute of Science and Technology,
50 UNIST-gil, Ulsan 44919, South Korea
}

\begin{abstract}
Brane tilings are bipartite graphs on a 2-torus 
that encode the Lagrangians of $4d$ $\mathcal{N}=1$ supersymmetric gauge theories
arising on D3-branes probing toric Calabi-Yau 3-folds.
Among these bipartite graphs, only those that satisfy geometric consistency conditions
correspond to well-behaved quantum field theories.
We train a convolutional neural network (CNN)
to distinguish geometrically consistent from inconsistent brane tilings
directly from their Kasteleyn matrices.
We study a family of $4d$ $\mathcal{N}=1$ theories
obtained by adding diagonal edges to the hexagonal faces of the brane tiling
for the abelian orbifold $\mathbb{C}^3/\mathbb{Z}_3\times\mathbb{Z}_3$,
and find that the CNN identifies geometric inconsistency with high accuracy.
For inconsistent brane tilings that can be rendered consistent by Higgsing a single
bifundamental chiral field,
we show that gradient-based saliency analysis can be used to localize 
the responsible chiral fields with accuracy well above a matched random baseline.
These results demonstrate that explainable AI (XAI) can be used to identify local defects
responsible for inconsistencies in supersymmetric gauge theories. 
\end{abstract}

\maketitle

\section{Introduction}

$4d$ $\mathcal{N}=1$ supersymmetric gauge theories arising 
in string theory provide a particularly rich setting for studying 
non-perturbative phenomena, including \textit{Seiberg duality} \cite{Seiberg:1994pq}. 
A fundamental question is which
apparently well-defined gauge theories correspond to 
consistent quantum field theories and admit a realization in string theory. 
Potential obstructions include gauge anomalies, 
apparent violations of the superconformal 
unitarity bound,
and vacuum moduli spaces that fail to reproduce the expected 
Calabi-Yau geometry. 
Diagnosing and classifying these obstructions is therefore essential 
for exploring the landscape of consistent $4d$ $\mathcal{N}=1$ theories.

Machine learning methods have been applied to a wide range of problems in 
string theory and mathematical physics \cite{He:2017aed, Krefl:2017yox, Ruehle:2017mzq, Carifio:2017bov, He:2017set, Hashimoto:2018ftp, Cole:2018emh, Halverson:2019tkf, Hashimoto:2019bih, Cole:2019enn, Jejjala:2019kio, Bies:2020gvf, Ruehle:2020jrk, Bao:2020nbi, Akutagawa:2020yeo, Gukov:2020qaj, Jejjala:2020wcc, Cole:2021nnt, Larfors:2021pbb, Abel:2021rrj, Berglund:2021ztg, Ri:2023xcn, Gukov:2024opc, Robinson:2026qsp}, including studies of Calabi-Yau geometries 
and their associated supersymmetric gauge theories \cite{Seong:2023njx, Choi:2023rqg, Seong:2024wkt, Heckman:2026xsi}. 
Although these approaches have demonstrated the ability of machine learning to classify and 
predict properties of complex physical and mathematical systems, 
the use of \textit{explainable AI} \cite{2013arXiv1312.6034S, 2015arXiv151204150Z, 2017arXiv170507874L, 2016arXiv161002391S, 2017arXiv171100399W, 2018arXiv181003292A, 2018arXiv180610758H} to uncover the physical origin of inconsistencies in quantum field theories 
remains largely unexplored.

In this work, we introduce a machine learning framework that not only identifies 
inconsistencies in a given $4d$ $\mathcal{N}=1$ supersymmetric gauge theory, 
but also locates the chiral field
that is most likely responsible for the inconsistency. 
We then show that giving the identified chiral field a non-zero vacuum expectation value (VEV) 
triggers Higgsing and, after the resulting massive fields are integrated out, yields a geometrically consistent
$4d$ $\mathcal{N}=1$ theory.
Our approach thus constitutes an application of explainable AI 
to supersymmetric gauge theories realized in string theory, 
with the model detecting defects in gauge theories, 
analogous to localizing pathological regions in medical images \cite{HAVAEI201718, 2017arXiv170204595Z, QUELLEC2017178, 10.3389/fnagi.2019.00194, SAYRES2019552}.

We focus on a family of $4d$ $\mathcal{N}=1$ 
supersymmetric gauge theories realized as worldvolume theories 
on a stack of D3-branes probing a toric Calabi-Yau 3-fold \cite{Douglas:1997de, Douglas:1996sw, Feng:2000mi, Feng:2001xr}. 
These theories admit a Type IIB brane configuration 
given in terms of a bipartite periodic graph on a 2-torus $T^2$, 
known as a \textit{brane tiling} \cite{Franco:2005rj, Hanany:2005ve, Franco:2005sm, 2003math.....10326K, Hanany:2012hi}. 
The Lagrangian of the corresponding $4d$ $\mathcal{N}=1$ theory 
is fully encoded in the brane tiling, 
with edges and faces of the graph corresponding to 
bifundamental chiral fields $X_{ij}$ and $U(N)_i$ gauge groups, 
respectively. 
The positive and negative terms of the superpotential $W$ 
are formed by gauge-invariant products of chiral fields 
that are associated with the white and black nodes 
of the bipartite graph, respectively.
Because every edge is incident on one node of each color, 
every associated chiral field appears exactly once 
in a positive and once in a negative term of the superpotential, 
which we refer to as the \textit{toric condition} \cite{Feng:2000mi, Feng:2001xr, Feng:2002zw}. 
This condition ensures that the F-terms are binomial 
and form a binomial ideal.
For a single D3-brane, for which all gauge factors are $U(1)_i$, 
the mesonic moduli space \cite{Butti:2007jv, Forcella:2008bb, Forcella:2008eh} is therefore toric \cite{fulton1993introduction, cox2011toric} and, 
for a consistent brane tiling, reproduces the toric Calabi-Yau 3-fold probed by the D3-brane.

Brane tilings provide a natural geometric and combinatorial framework for determining whether
the corresponding $4d$ $\mathcal{N}=1$ theory is consistent.
Among the various known consistency conditions \cite{Hanany:2005ss, Gulotta:2008ef, 2009arXiv0901.4662B, 2011arXiv1104.1592B, Hanany:2015tgh}, 
brane tilings identify geometrically whether the corresponding $4d$ $\mathcal{N}=1$ theory 
admits a non-degenerate positive assignment of $U(1)_R$-charges \cite{Intriligator:2003jj, Butti:2005vn, Martelli:2005tp, Martelli:2006yb, Gulotta:2008ef, Hanany:2011bs} to the chiral fields.
Such consistent $U(1)_R$-charges satisfy, 
\beal{es01a01}
\sum_{X_{ij} \in W_k} R(X_{ij}) = 2 ~,~
\eea
for every superpotential term $W_k$ represented by either a white or black node, and 
\beal{es01a02}
\sum_{X_{ij} \in F_i} (1-R(X_{ij})) = 2 ~,~
\eea
for every face $F_i$ representing a gauge group in the brane tiling.
Here, \eref{es01a01} ensures that every superpotential term has $U(1)_R$-charge 2, while \eref{es01a02}
ensures the vanishing of the NSVZ beta function for each gauge group at the superconformal fixed point.
We refer to a brane tiling and its corresponding $4d$ $\mathcal{N}=1$ theory as \textit{geometrically consistent} \cite{Hanany:2005ss, Gulotta:2008ef, 2009arXiv0901.4662B, 2011arXiv1104.1592B, Hanany:2015tgh}
when it satisfies the above conditions. 

Geometric consistency of a brane tiling can be characterized by the existence of an isoradial embedding \cite{Hanany:2005ss}
and by properly ordered, non-self-intersecting zig-zag paths \cite{2009arXiv0901.4662B, 2011arXiv1104.1592B}.
In this work, we make use of the equivalent condition 
for the number of gauge groups $N_G$
to be twice the area $A(\Delta)$ of the lattice polygon in $\mathbb{Z}^2$ representing the toric diagram $\Delta$ of the corresponding toric Calabi-Yau 3-fold, 
\beal{es01a03}
2 A(\Delta) = N_G~,~
\eea
in order for the brane tiling to be geometrically consistent. 

While these conditions provide powerful global diagnostics for determining whether a brane tiling is geometrically consistent, 
they do not identify which component of the bipartite graph or the Lagrangian of the corresponding $4d$ $\mathcal{N}=1$ theory 
is responsible for the inconsistency, nor how the theory can be modified to restore geometric consistency.  
Using a dataset of consistent and inconsistent brane tilings, we show that our proposed machine learning model accurately 
distinguishes consistent $4d$ $\mathcal{N}=1$ theories from inconsistent ones.
More importantly, our machine learning model identifies through explainable AI techniques 
individual chiral fields of the $4d$ $\mathcal{N}=1$ theory which, 
through Higgsing, can be removed in order to yield
geometrically consistent brane tilings and $4d$ $\mathcal{N}=1$ theories.
These results establish a direct connection between explainable machine learning and diagnostics for physical consistency in quantum field theories, 
providing a machine learning framework in which AI not only detects inconsistencies in supersymmetric gauge theories, but also suggests how they can be resolved.
\\

\section{Background}

\subsection{Brane Tilings and the Kasteleyn Matrix}

\begin{figure}[htt!!]
\begin{center}
\resizebox{1\hsize}{!}{
\includegraphics[height=5cm]{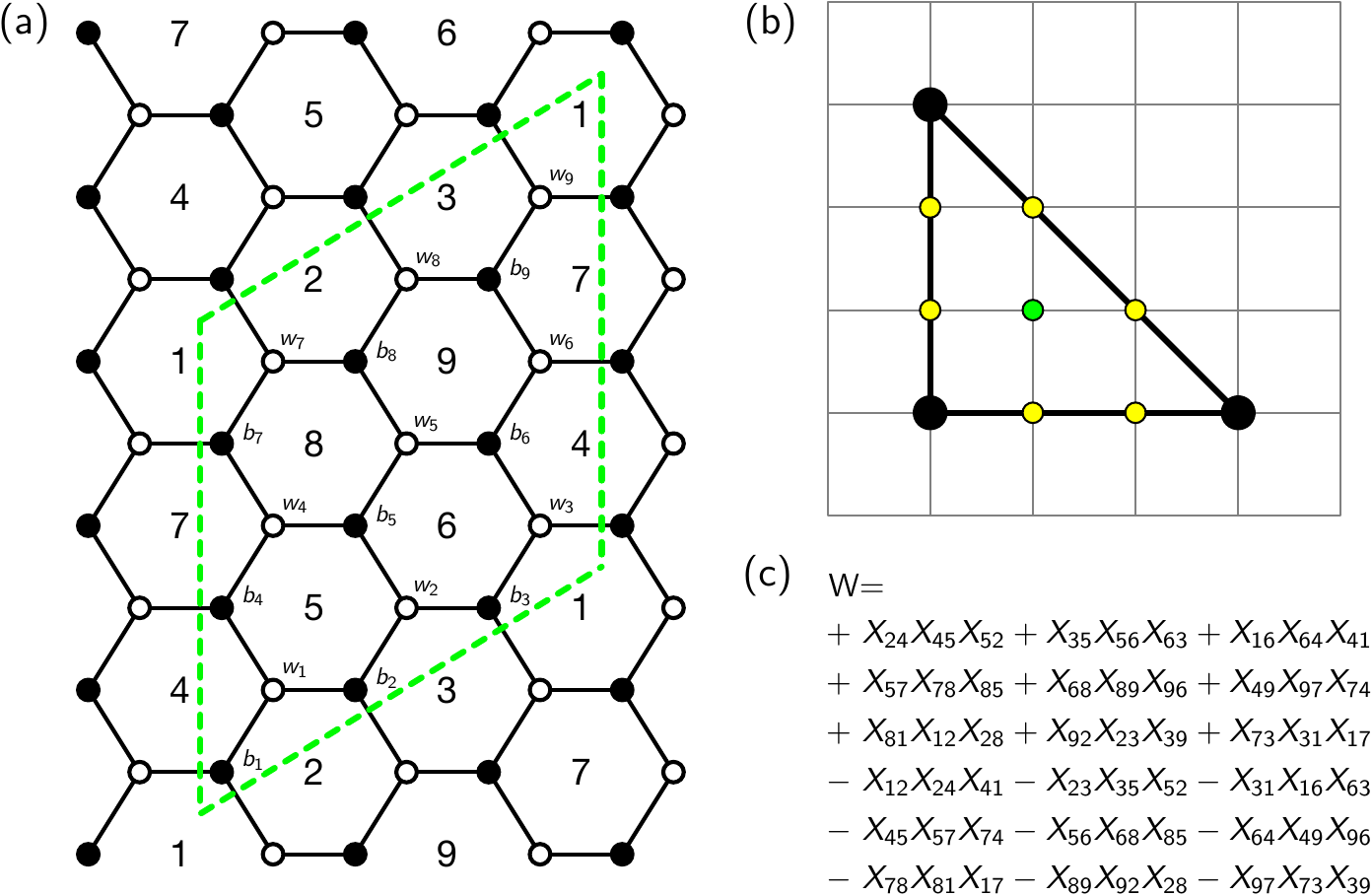} 
}
\caption{
(a) Brane tiling for $\mathbb{C}^3/\mathbb{Z}_3 \times \mathbb{Z}_3$ with orbifold action $(0,1,2)(1,0,2)$. 
The outlined region (green) indicates the chosen fundamental domain on the 2-torus $T^2$.
The corresponding (b) toric diagram
and (c) superpotential $W$ of the associated $4d$ $\mathcal{N}=1$ theory are also shown.
\label{fig_01}}
 \end{center}
 \end{figure}

For a single probe D3-brane, brane tilings and the corresponding 
$4d$ $\mathcal{N}=1$ supersymmetric gauge theories have abelian gauge groups
and their mesonic moduli spaces \cite{Butti:2007jv, Forcella:2008bb, Forcella:2008eh} are toric Calabi-Yau 3-folds.
The geometry of the toric variety is given by a convex lattice polygon $\Delta$ in $\mathbb{Z}^2$, 
which we refer to as the \textit{toric diagram} \cite{fulton1993introduction, cox2011toric}.
 The toric diagram $\Delta$ can be considered as the Newton polygon of the following \textit{Newton polynomial}, 
 \beal{es02a01}
 P(x,y) = \sum_{(n_x,n_y) \in \Delta} c_{(n_x,n_y)} x^{n_x} y^{n_y} ~,~
 \eea
 where $x,y \in \mathbb{C}^*$,
 and $(n_x,n_y) \in \Delta \subset \mathbb{Z}^2$ correspond to the coordinates of
 vertices in the toric diagram $\Delta$.
We refer to the complex coefficients $c_{(n_x,n_y)} \in \mathbb{C}^*$ 
as the complex structure moduli of the mirror Calabi-Yau 3-fold \cite{Feng:2005gw, Hanany:2011bs, Hori:2000kt}.
When the edge weights are set to unity, 
the absolute value of the coefficient associated with vertices in $\Delta$
gives the multiplicity of the corresponding \textit{perfect matchings} \cite{Kenyon:2003uj, Hanany:2006nm}.
 
The perfect matchings and
the Newton polynomial can be constructed directly from the bipartite graph of the brane tiling through
its weighted adjacency matrix known as the \textit{Kasteleyn matrix} $K(x,y)$ \cite{kasteleyn1967graph}.
The rows and columns of $K(x,y)$
correspond to white nodes $w_u$ and 
black nodes $b_v$ of the brane tiling.
The entry $K_{uv}$ identifies the edges connecting $w_u$ and $b_v$
in terms of a 
monomial $x^{n_x} y^{n_y}$,
where the exponents $(n_x,n_y) \in \mathbb{Z}^2$ determine how the edges
cross the boundaries of the fundamental domain of the brane tiling on the 2-torus $T^2$ as shown in \fref{fig_01}.
We use the same fixed ordering of the white and black nodes for every
brane tiling, inherited from the labeling in \fref{fig_01}(a).
For multiple distinct edges between the same pair of nodes, 
the monomial contributions are summed up in $K_{uv}$.
The permanent of the Kasteleyn matrix enumerates the perfect matchings
and yields the Newton polynomial in \eref{es02a01},
\beal{es02a01b}
\text{perm}~ K(x,y) = P(x,y) ~.~
\eea

\subsection{Hexagonal Brane Tilings with Diagonals}

\begin{figure}[htt!!]
\begin{center}
\resizebox{0.8\hsize}{!}{
\includegraphics[height=5cm]{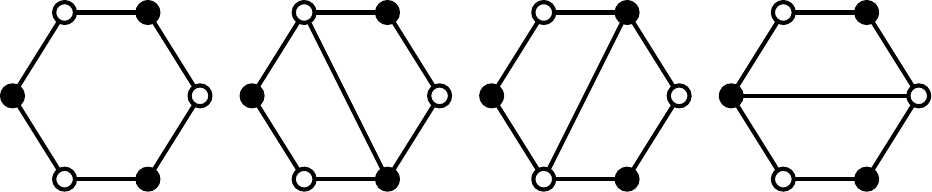} 
}
\caption{
A hexagonal face and the $3$ possible orientations of an added diagonal edge
in a brane tiling. 
\label{fig_02}}
 \end{center}
 \end{figure}

Brane tilings consisting of hexagonal faces
correspond to abelian orbifolds of $\mathbb{C}^3$ \cite{Davey:2010px, Hanany:2010ne, Hanany:2010cx}, where the order of the finite abelian quotienting group $\Gamma \subset SU(3)$
is the number of distinct faces in the hexagonal brane tiling.
\fref{fig_01}(a) illustrates the hexagonal brane tiling for the abelian orbifold of the form 
$\mathbb{C}^3/\mathbb{Z}_3 \times \mathbb{Z}_3$ with orbifold action $(0,1,2)(1,0,2)$.
The corresponding superpotential is shown in \fref{fig_01}(c), 
where the positive and negative cubic terms correspond to degree 3 white and black nodes in the brane tiling, respectively.

Based on the fundamental domain shown in \fref{fig_01}(a), 
the brane tiling for $\mathbb{C}^3/\mathbb{Z}_3 \times \mathbb{Z}_3$ $(0,1,2)(1,0,2)$
has the following Kasteleyn matrix,
\beal{es02b02}
K(x,y) =
\resizebox{0.72\linewidth}{!}{$
\left(
\ba{c|ccccccccc}
\; & b_1 & b_2 & b_3 & b_4 & b_5 & b_6 & b_7 & b_8 & b_9 \\
\hline
w_1 & 1 & 1 & 0 & 1 & 0 & 0 & 0 & 0 & 0 \\
w_2 & 0 & 1 & 1 & 0 & 1 & 0 & 0 & 0 & 0 \\
w_3 & x & 0 & 1 & 0 & 0 & 1 & 0 & 0 & 0 \\
w_4 & 0 & 0 & 0 & 1 & 1 & 0 & 1 & 0 & 0 \\
w_5 & 0 & 0 & 0 & 0 & 1 & 1 & 0 & 1 & 0 \\
w_6 & 0 & 0 & 0 & x & 0 & 1 & 0 & 0 & 1 \\
w_7 & y & 0 & 0 & 0 & 0 & 0 & 1 & 1 & 0 \\
w_8 & 0 & y & 0 & 0 & 0 & 0 & 0 & 1 & 1 \\
w_9 & 0 & 0 & y & 0 & 0 & 0 & x & 0 & 1
\ea
\right)
$}
~,~
\nn\\
\eea
where the permanent gives,
\beal{es02b03}
&&
P(x,y) = 1 + x^3 + y^3 + 21  x y
\nn\\
&&
\hspace{1.2cm}
+ 3 \left( x + y + x^2 + y^2 + x^2 y + x y^2 \right)
~,~
\eea
whose Newton polygon is the lattice triangle with extremal vertices
$(0,0)$, $(3,0)$ and $(0,3)$ as illustrated in \fref{fig_01}(b).
This is the toric diagram for $\mathbb{C}^3/\mathbb{Z}_3 \times \mathbb{Z}_3$ $(0,1,2)(1,0,2)$.

Each hexagonal face of the brane tiling in \fref{fig_01}(a)
can be divided into two quadrilateral faces by adding a \textit{diagonal edge}
in 3 different orientations as illustrated in \fref{fig_02}.
Each added diagonal edge introduces an additional bifundamental chiral field
and an additional $U(N)$ gauge group to the associated $4d$ $\mathcal{N}=1$ gauge theory.
While the number of superpotential terms remains $N_W=18$,
the two superpotential terms associated with the end-nodes of the added diagonal edge
each gain the new corresponding chiral field.
As a result, a hexagonal brane tiling with $N_D$ diagonal edges has
the following number of faces and edges,
\beal{es02b04}
N_G = 9 + N_D ~,~ \quad N_E = 27 + N_D ~,~
\eea
respectively.

As discussed in \eref{es01a03},
for such a hexagonal brane tiling with diagonals to be consistent,
the corresponding toric diagram obtained from the Kasteleyn matrix
as given in \eref{es02a01b} has to have an area satisfying $2 A = 9+ N_D$.
Allowing each of the $9$ faces either to have no diagonal or to contain one of the $3$ possible diagonals
in the brane tiling for $\mathbb{C}^3/\mathbb{Z}_3 \times \mathbb{Z}_3$ $(0,1,2)(1,0,2)$
yields a set of $4^9 = 262{,}144$ brane tilings.
Of these, we identify $1{,}387$ to be consistent brane tilings
while the remaining $260{,}757$ brane tilings are inconsistent.
\\

\section{Machine Learning Framework}
\label{section_cnn}

\subsection{Dataset and Brane Tiling Representation}

In this paper, we would first like to make use of a convolutional neural network (CNN) \cite{6795724, NIPS1989_53c3bce6, 726791}
in order to identify whether a given hexagonal brane tiling with diagonals
and its corresponding $4d$ $\mathcal{N}=1$ supersymmetric gauge theory are geometrically consistent.
In order to make use of a CNN, 
we convert the Kasteleyn matrix of a brane tiling
into a tensor suitable for CNN input. 
The hexagonal brane tiling
for $\mathbb{C}^3/\mathbb{Z}_3 \times \mathbb{Z}_3$ with orbifold action $(0,1,2)(1,0,2)$
and all brane tilings obtained from it by adding diagonal edges to the hexagonal faces 
have Kasteleyn matrices $K(x,y)$
with entries $K_{uv}$, where $u,v = 1, \dots, 9$.
For the fundamental domain shown in \fref{fig_01}(a),
all exponents of $x$ and $y$ satisfy $n_x, n_y \in \{-1,0,1\}$
including those associated with the added diagonal edges.

The Kasteleyn matrix $K(x,y)$
is expressed in terms of a 
CNN input tensor $\mathbf{x} \in \mathbb{R}^{9\times 9 \times 9}$, which we refer to as the \textit{Kasteleyn tensor},
where each polynomial entry $K_{uv}$
is expressed in terms of a 9-dimensional channel $\mathbf{x}_{uv}$.
Here, 
each monomial in $K_{uv}$ takes the form $m_{(n_x,n_y)}x^{n_x}y^{n_y}$,
where $m_{(n_x,n_y)}$ is taken to be a non-negative integer.
The component $\mathbf{x}_{uvk}$ takes the value of $m_{(n_x,n_y)}$
when the corresponding monomial is present in the Kasteleyn matrix
and vanishes otherwise.
The index $k$ is given by,
\beal{es03a01}
k = 3(n_x + 1) + (n_y + 1) ~,~ \quad k\in\{0,1,\dots,8\} ~.~
\eea

For each input tensor $\mathbf{x}$,
we assign a binary consistency label $y\in \{0,1\}$.
Among the $262{,}144$ brane tilings, 
$260{,}757$ are geometrically inconsistent and are assigned $y=0$, 
whereas the remaining $1{,}387$ are consistent and are assigned $y=1$.
We note here that geometric consistency is a global property of a brane tiling,
whereas the input tensor $\mathbf{x}$ encodes information associated with individual entries and monomials of the Kasteleyn matrix. 
The challenge for our CNN is therefore to identify global consistency based on local properties encoded in the brane tilings in our database.

\subsection{CNN Architecture}

\begin{figure*}[htt!!]
\begin{center}
\resizebox{1\hsize}{!}{
\includegraphics[height=5cm]{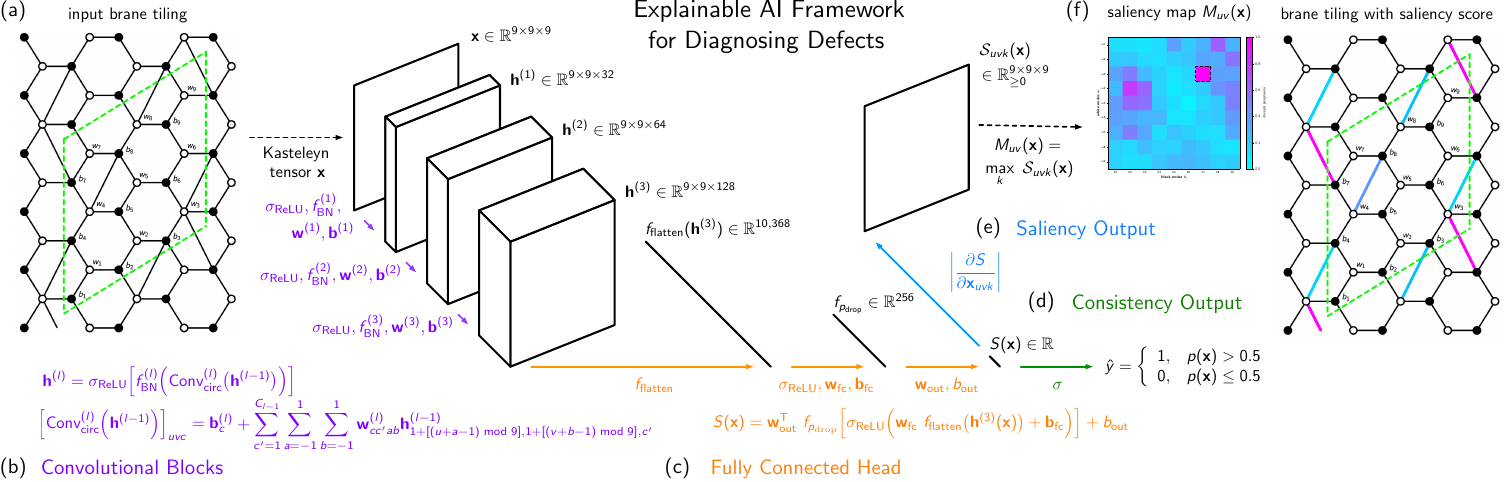} 
}
\caption{
Explainable-AI framework for identifying and locating defects in brane tilings and $4d$ $\mathcal{N}=1$ supersymmetric gauge theories.
(a) An input brane tiling is encoded by its Kasteleyn tensor $\mathbf{x}$,
which becomes the input for
(b) $3$ circularly padded convolutional blocks 
that map $\mathbf{x}$ to feature tensors with $32$, $64$, and $128$ channels while preserving the $9\times9$ spatial resolution.
(c) The final feature tensor is flattened and passed through a fully connected head, a $256$-unit layer followed by a linear output layer, to produce the logit $S(\mathbf{x})$.
(d) A sigmoid function and decision threshold of $p_c=0.5$ determine the predicted consistency label $\hat{y}(\mathbf{x})$.
(e) The input gradient of the logit defines the saliency tensor $\mathcal{S}_{uvk}(\mathbf{x})$.
(f) Maximization over the channel index produces the projected saliency
map $M_{uv}(\mathbf{x})$, which is used to visualize saliency as a heat
map and as diagonal edge colors on the input brane tiling.
Quantitative localization of defects is evaluated using the components of the saliency tensor
$\mathcal{S}_{uvk}(\mathbf{x})$.
\label{fig_03}}
 \end{center}
 \end{figure*}

As illustrated in \fref{fig_03},
our CNN consists of 3 successive 2-dimensional convolutional blocks
with 32, 64 and 128 filters, respectively.
Each block is of the form, 
\beal{es04a000}
\mathbf{h}^{(l)}
= \sigma_{\text{ReLU}}
\Big[
f_{\text{BN}}^{(l)}\Big(
\text{Conv}^{(l)}_{\text{circ}}\big(
\mathbf{h}^{(l-1)}
\big)
\Big)
\Big]~,~
\eea
and contains a $3\times 3$ convolution,
\beal{es04a000b}
&&
\Big[
\text{Conv}_{\text{circ}}^{(l)}
\Big(
\mathbf{h}^{(l-1)}
\Big)
\Big]_{uvc}
=
\mathbf{b}_c^{(l)}
+ \sum_{c^\prime = 1}^{C_{l-1}} \sum_{a=-1}^{1} \sum_{b=-1}^{1}
\nn\\
&&
\hspace{0.5cm}
\times 
\mathbf{w}_{cc^\prime a b}^{(l)} 
\mathbf{h}^{(l-1)}_{
1+[(u+a-1)\bmod{9}],
1+[(v+b-1)\bmod{9}],
c^\prime
}~,~
\nn\\
\eea
where $u,v=1, \dots, 9$, 
$c=1, \dots, C_l$, 
$c^\prime=1, \dots, C_{l-1}$,
and $a,b \in \{-1,0,1\}$ with $(C_0, C_1, C_2, C_3)=(9,32,64,128)$.
Here, we have circular padding in order to avoid the introduction of distinguished boundary values in the input Kasteleyn tensor $\mathbf{h}^{(0)}=\mathbf{x}$.
Moreover, each block contains batch normalization $f_{\text{BN}}^{(l)}$ 
and a rectified linear unit (ReLU) activation $\sigma_{\text{ReLU}}$.
We deliberately omit spatial pooling in order to preserve the full $9 \times 9$ matrix resolution, 
which we later make use of in order to 
trace the origin of a predicted inconsistency back to individual entries
$K_{uv}$ of an input brane tiling.

The final convolutional block produces a $9 \times 9 \times 128$ feature tensor, 
which is flattened into a $10{,}368$-dimensional vector and passed through a fully connected layer with $256$ neurons.
This is then followed by a ReLU activation $\sigma_{\text{ReLU}}$ and dropout $f_{p_{\mathrm{drop}}}$ \cite{srivastava2014dropout}
with probability $p_{\mathrm{drop}}=0.6$.
A final linear layer maps the resulting representation to the scalar logit given by, 
\beal{es04a00}
&&
S(\mathbf{x})
= \mathbf{w}_{\text{out}}^{\mathsf{T}} 
~f_{p_{\mathrm{drop}}}
\Big[
\sigma_{\text{ReLU}} \Big(
\mathbf{w}_{\text{fc}} ~f_\text{flatten}\big(
\mathbf{h}^{(3)}(\mathbf{x})
\big)
\nn\\
&&
\hspace{4cm}
+\mathbf{b}_{\text{fc}}
\Big)
\Big] + b_{\text{out}}
~,~
\eea
where $\mathbf{h}^{(3)}(\mathbf{x}) \in \mathbb{R}^{9 \times 9 \times 128}$ is the output of the third convolutional block, 
$f_\text{flatten}$ denotes the flattening of output tensor,
$\mathbf{w}_{\text{fc}} \in \mathbb{R}^{256 \times 10{,}368}$ and $\mathbf{b}_{\text{fc}} \in \mathbb{R}^{256}$ are the parameters of the 256-unit fully connected layer,
and $\mathbf{w}_{out} \in \mathbb{R}^{256}$ and $b_{out} \in \mathbb{R}$ are the parameters of the final linear layer.
 
The probability that the brane tiling represented by $\mathbf{x}$ is
geometrically consistent is obtained by the following sigmoid function,
\beal{es04a01}
p(\mathbf{x})
=
\sigma\!\left(S(\mathbf{x})\right)
=
\frac{1}{1+\exp[-S(\mathbf{x})]}
~.~
\eea
Accordingly, the CNN maps the information encoded in the input Kasteleyn tensor $\mathbf{x}$ to
the global predicted consistency label $\hat{y}\in\{0,1\}$, where 
brane tilings with
$p(\mathbf{x})> p_c= 0.5$ are assigned $\hat{y}(\mathbf{x}) = 1$ and classified as geometrically consistent, 
whereas those with $p(\mathbf{x})\leq p_c = 0.5$ are assigned $\hat{y}(\mathbf{x}) = 0$ and classified as inconsistent.

\begin{table}[t]
\centering
\begin{tabular*}{\linewidth}{@{\extracolsep{\fill}}lr@{}}
\hline
\text{Metric} & \text{Value} \\
\hline
Accuracy & 99.92\% \\
Precision $P$ & 87.72\% \\
Recall $R$ & 98.41\% \\
$F_1$-score & 92.76\% \\
ROC-AUC & 99.99\% \\
\hline
Consistent brane tilings recovered & 1,365 of 1,387 \\
False negatives FN & 22 \\
False positives FP & 191 \\
\hline
\end{tabular*}
\caption{
Classification performance of the trained CNN on the complete set of
$262{,}144$ brane tilings at the decision threshold $p_c =0.5$.
Precision ($P$), recall ($R$), and the $F_1$-score refer to the geometrically consistent brane tilings. 
The ROC-AUC \cite{FAWCETT2006861} is threshold independent.
\label{tab_00}
}
\end{table}

\subsection{Training}

Training and validation of the CNN is done using 
all $1{,}387$ geometrically consistent brane tilings 
together with $13{,}870$ inconsistent ones, which gives approximately a $1{:}10$ ratio between the two classes.
The inconsistent brane tilings are sampled according to their difficulty, 
measured by the deficit $N_G - 2 A$, 
rather than uniformly, in order to prevent the negative subset from being dominated by easily classified inconsistent brane tilings.
The CNN parameters are optimized using the Adam optimizer
and the binary cross-entropy loss, which is for a training batch of $N_S$ samples, given in our case by,
\beal{es04a10}
\mathcal{L}
=
-\frac{1}{N_S}
\sum_{n=1}^{N_S}
\left[
y_n\log p_n
+
(1-y_n)\log(1-p_n)
\right]
~,~
\eea
where 
$y_n\in \{0,1\}$ is the consistency label of the $n$-th brane tiling sample.
The cross-entropy loss is evaluated directly from the probability $p_n=p(\mathbf{x}_n)$, which is the predicted probability for the $n$-th brane tiling to be geometrically consistent as defined in \eref{es04a01}.

Because geometrically consistent brane tilings
are rare in the dataset, we make use of the $F_1$-score as the primary validation metric 
rather than the overall classification accuracy.
The precision $P$ and recall $R$ are defined by,
\beal{es04a15}
P
=
\frac{\mathrm{TP}}{\mathrm{TP}+\mathrm{FP}}
~,~
R
=
\frac{\mathrm{TP}}{\mathrm{TP}+\mathrm{FN}}
~,~
\eea
and
the $F_1$-score takes the form, 
\beal{es04a16}
F_1
=
\frac{2~ P R}{P+R}
=
\frac{
2~ \mathrm{TP}
}{
2 ~\mathrm{TP}+\mathrm{FP}+\mathrm{FN}
}
~.~
\eea
Here, 
$\mathrm{TP}$, $\mathrm{FP}$, and $\mathrm{FN}$ denote the numbers
of true positives, false positives, and false negatives, respectively,
where geometrically consistent brane tilings form the positive class.
The $F_1$-score is sensitive to both consistent
brane tilings that are missed by the classifier and to inconsistent brane tilings
that are incorrectly classified as consistent, 
which makes it a more informative measure for validating the CNN and selecting its parameters.

\subsection{Explainable AI and Localization of Inconsistency}

We make use of \textit{gradient-based saliency} \cite{2013arXiv1312.6034S}
in order to identify which components of the Kasteleyn matrix input tensor $\mathbf{x}$
and hence which part of the corresponding brane tiling and $4d$ $\mathcal{N}=1$ supersymmetric gauge theory
most strongly influence the CNN's prediction of geometric inconsistency.
We recall $S(\mathbf{x})$ as the output logit of the trained CNN for an input Kasteleyn tensor $\mathbf{x}$.
For a small perturbation of the input tensor $\delta \mathbf{x}$, the output logit admits a first-order expansion
of the following form, 
\beal{es05a01}
S( \mathbf{x} + \delta \mathbf{x}) = 
S(\mathbf{x}) + \sum_{u,v,k} \frac{\partial S(\mathbf{x})}{\partial \mathbf{x}_{uvk}} ~\delta \mathbf{x}_{uvk} + 
\mathcal{O}(\|\delta\mathbf{x}\|^2)
~,~\nn\\
\eea
where the first order sensitivity of the output logit 
is given by the following saliency tensor, 
\beal{es05a10}
\mathcal{S}_{uvk}(\mathbf{x}) = \left|
\frac{\partial S(\mathbf{x})}{\partial \mathbf{x}_{uvk}} 
\right|
~.~
\eea
Here, 
$\mathcal{S}(\mathbf{x})
=\{\mathcal{S}_{uvk}(\mathbf{x})\}
\in\mathbb{R}^{9\times9\times9}_{\geq0}$,
$k=0,\dots, 8$ and $(u,v)$ labels the white and black node pair $(w_u, b_v)$ in the input brane tiling.

The saliency tensor can be projected to a $9\times 9$ \textit{saliency map} \cite{2013arXiv1312.6034S} given by, 
\beal{es05a15}
M_{uv} (\mathbf{x}) = 
\max_k ~\mathcal{S}_{uvk}(\mathbf{x}) ~.~
\eea
For visualization, we normalize $M_{uv}(\mathbf{x})$ such that its values lie in $[0,1]$.
Larger values of $M_{uv}(\mathbf{x})$ highlight node pairs whose associated Kasteleyn matrix entries have greater local sensitivity of the CNN output logit.
Because of the projection in \eref{es05a15}, the global maximum of the saliency map occurs at the same node pair as the global maximum of the saliency tensor.
When the ranking is restricted to diagonal edges, however, the two are not necessarily equivalent,
since $M_{uv}(\mathbf{x})$ retains the maximum over all channels $k$ at a node pair, including components that carry no edge.
Since the projection to $M_{uv}(\mathbf{x})$ discards the channel index $k$,
we use $M_{uv}(\mathbf{x})$ only for visualization, 
while all localization accuracies are computed by ranking the $9\times 9 \times 9=729$ components $\mathcal{S}_{uvk}(\mathbf{x})$ directly.
When indicated, this ranking is restricted to components corresponding to diagonal edges actually present in the input brane tiling.

\begin{figure}[htt!!]
\begin{center}
\resizebox{1\hsize}{!}{
\includegraphics[height=5cm]{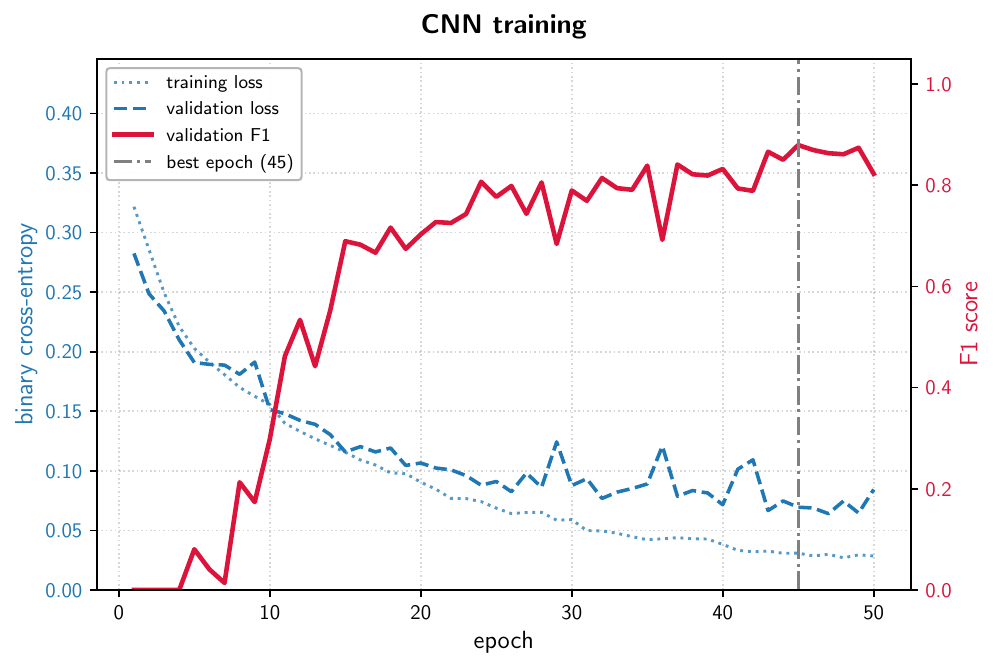} 
}
\caption{
Training and validation binary cross-entropy losses 
and validation $F_1$-score over $50$ epochs. 
The $F_1$-score is evaluated at $p_c=0.5$. 
The vertical line marks the selected checkpoint at epoch $45$, which maximizes the validation $F_1$-score.
\label{fig_04}}
 \end{center}
 \end{figure}

\section{Results}

\subsection{Training Performance}

The training pool of
$1{,}387$ consistent and $13{,}870$ inconsistent brane tilings
is partitioned into disjoint training and validation sets using a stratified $9{:}1$ split, 
which approximately preserves the $1{:}10$ class ratio 
of consistent to inconsistent brane tilings in both sets.
This yields
a training set of $13{,}728$ brane tilings and a validation set of $1{,}529$ brane tilings.
The CNN is trained for $50$ epochs.
\fref{fig_04} shows the training and validation losses with the validation $F_1$-score
evaluated using a decision threshold of $p_c=0.5$ for the sigmoid output in \eref{es04a01}.
The losses decrease during training while the validation $F_1$-score improves
until it reaches its best value at epoch $45$.
For all subsequent analysis in this work, we
use the network parameters obtained at the epoch with the highest validation $F_1$-score.
If several epochs have the same $F_1$-score, we choose the one with the lowest validation loss.

The trained CNN is then used to identify consistency of the $262{,}144$ brane tilings in our dataset.
The performance of the trained CNN on the complete dataset is summarized in \tref{tab_00}.
The CNN classifier correctly identifies $1{,}365$ of the $1{,}387$ geometrically consistent brane tilings
corresponding to a $98.41\%$ recall rate.
Out of the brane tilings identified as geometrically consistent, $87.72\%$ are truly consistent, which gives an $F_1$-score of $92.76\%$.
Here, at the threshold $p_c=0.5$,
the corresponding $22$ false negatives and $191$ false positives give an overall accuracy of $99.92\%$.
The threshold-independent ROC-AUC is $99.99\%$ \cite{FAWCETT2006861}.

\begin{figure}[htt!!]
\begin{center}
\resizebox{0.95\hsize}{!}{
\includegraphics[height=5cm]{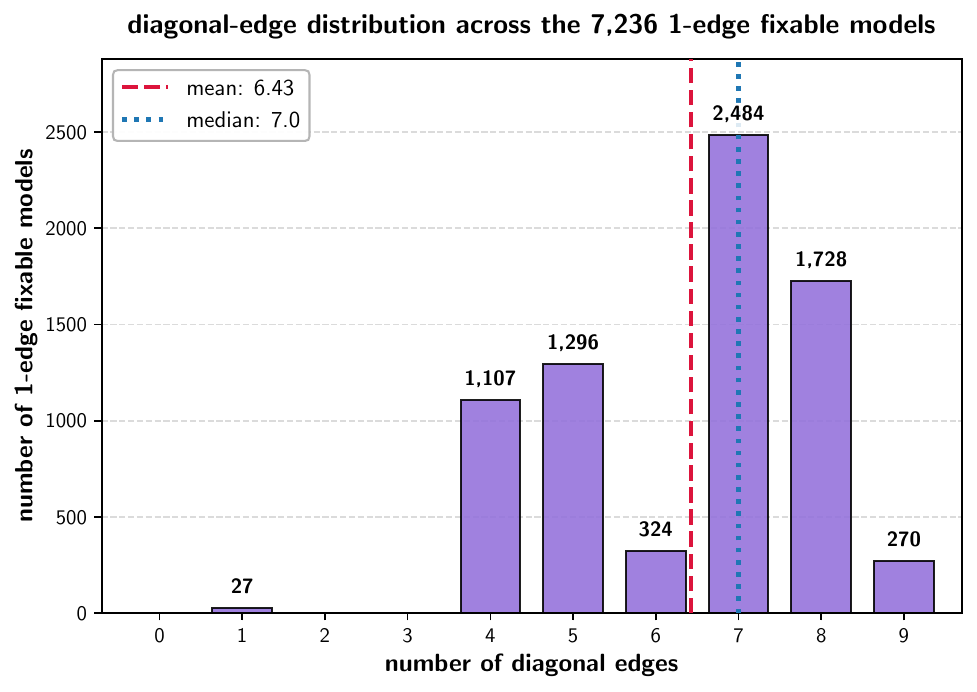} 
}
\caption{
Distribution of the number of diagonal edges $N_D$ among the $7{,}236$ $1$-edge-fixable brane tilings. 
The dashed (red) and dotted (blue) vertical lines indicate the mean $6.43$ and median $7$, respectively.
\label{fig_05}}
 \end{center}
 \end{figure}

We note that the overall high accuracy must be interpreted in the context of the strong imbalance between consistent and inconsistent brane tilings in the dataset.
Because $260{,}757$ of the $262{,}144$ brane tilings are inconsistent, 
a trivial classifier that labels every brane tiling as inconsistent would already achieve an accuracy of $99.47\%$.
Therefore, the $F_1$-score for identifying consistent brane tilings is substantially more informative than the overall accuracy.
This evaluation includes brane tilings used for training and validation and therefore characterizes only how the trained CNN partitions the complete brane tiling dataset.
It should therefore not be interpreted as a measure of out-of-sample generalization.

\begin{figure*}[htt!!]
\begin{center}
\resizebox{0.95\hsize}{!}{
\includegraphics[height=5cm]{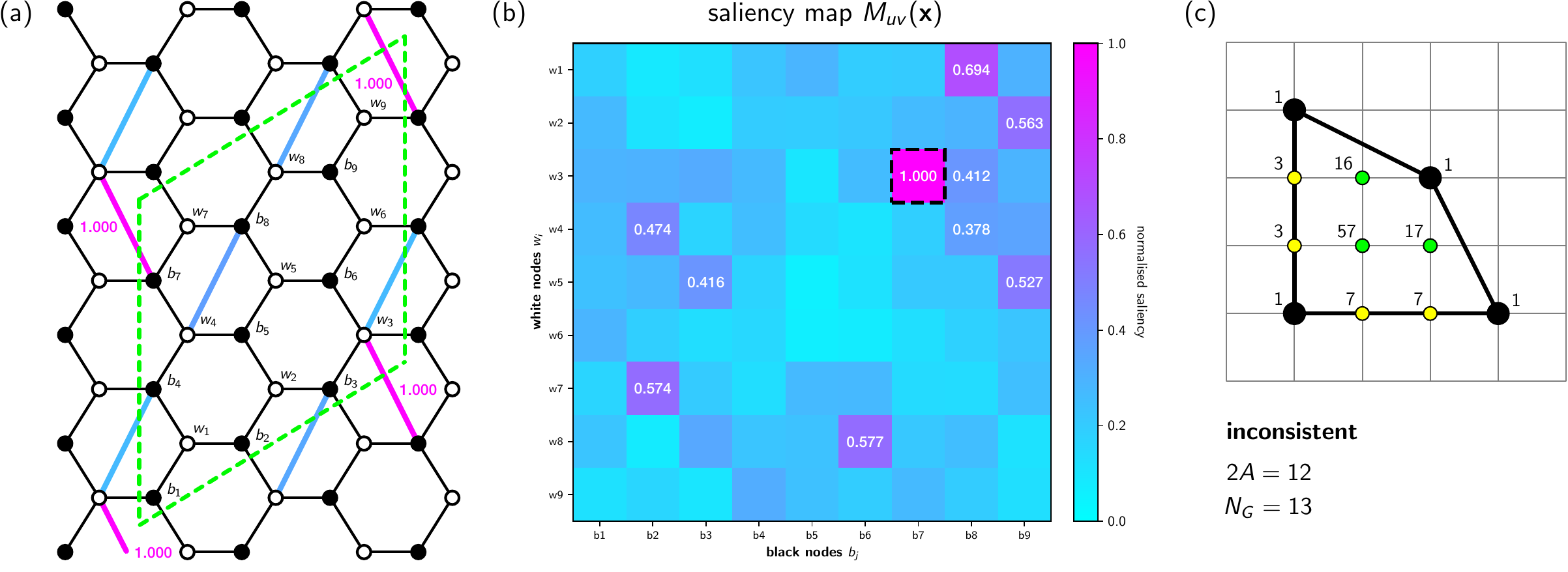} 
}
\caption{
Representative localization of a defect in a $1$-edge-fixable brane tiling.
(a)
Inconsistent brane tiling with coloring of the diagonal edges 
indicating normalized saliency from the saliency map $M_{uv}(\mathbf{x})$. 
The maximum saliency identifies the edge connecting $(w_3,b_7)$ in the brane tiling.
(b) Corresponding saliency map $M_{uv}(\mathbf{x})$ with $M_{37}(\mathbf{x})=1.000$ as highlighted.
(c) Toric diagram, for which $2A=12\neq N_G=13$, where the numbers on vertices denote multiplicities of the perfect matchings in the brane tiling.
Removing the diagonal edge between $(w_3,b_7)$ by Higgsing the corresponding bifundamental chiral field gives $N_G=12$ 
and restores consistency.
\label{fig_06}}
 \end{center}
 \end{figure*}

\subsection{Inconsistency Localization Performance}

\begin{table*}[t]
\centering
\begin{tabular*}{\textwidth}{@{\extracolsep{\fill}}ccccccc@{}}
\hline
\text{Number of} & \text{Total} & \text{Brane Tilings} & \text{Brane Tilings} & \text{Mean Defect} & \text{Top-1} & \text{Top-3} \\
\text{Diagonals} $N_D$ & \text{Brane Tilings} & \text{with $N_R=1$} & \text{with $N_R=2$} & \text{Fraction} $\langle N_R/N_D \rangle$ & \text{Accuracy} & \text{Accuracy} \\
\hline
1 & 27 & 27 & 0 & 100.00\% & 26 (96.30\%) & 27 (100\%) \\
4 & 1,107 & 972 & 135 & 28.05\% & 1,009 (91.15\%) & 1,102 (99.55\%) \\
5 & 1,296 & 1,080 & 216 & 23.33\% & 1,057 (81.56\%) & 1,146 (88.43\%) \\
6 & 324 & 324 & 0 & 16.67\% & 193 (59.57\%) & 226 (69.75\%) \\
7 & 2,484 & 1,512 & 972 & 19.88\% & 2,288 (92.11\%) & 2,471 (99.48\%) \\
8 & 1,728 & 1,080 & 648 & 17.19\% & 1,263 (73.09\%) & 1,513 (87.56\%) \\
9 & 270 & 162 & 108 & 15.56\% & 164 (60.74\%) & 192 (71.11\%) \\
\hline
\text{Total} & 7,236 & 5,157 & 2,079 & $21.10\%$ & \text{$6{,}000$ ($82.92\%$)} & \text{$6{,}677$ ($92.27\%$)} \\
\hline
\end{tabular*}
\caption{
Defect localization results for the $7{,}236$ $1$-edge-fixable brane
tilings, grouped by the number of diagonal edges $N_D$.
The Top-$1$ and Top-$3$ localization accuracies are obtained by ranking
all $729$ saliency tensor components $\mathcal{S}_{uvk}(\mathbf{x})$.
Here, $N_R$ is the number of diagonal edges whose individual removal restores consistency.
The mean defect fraction $\langle N_R/N_D\rangle$ measures the fraction of present diagonal edges 
whose removal restores consistency and serves as a descriptor 
of localization difficulty. 
\label{tab_01}
}
\end{table*}

\begin{figure*}[htt!!]
\begin{center}
\resizebox{0.95\hsize}{!}{
\includegraphics[height=5cm]{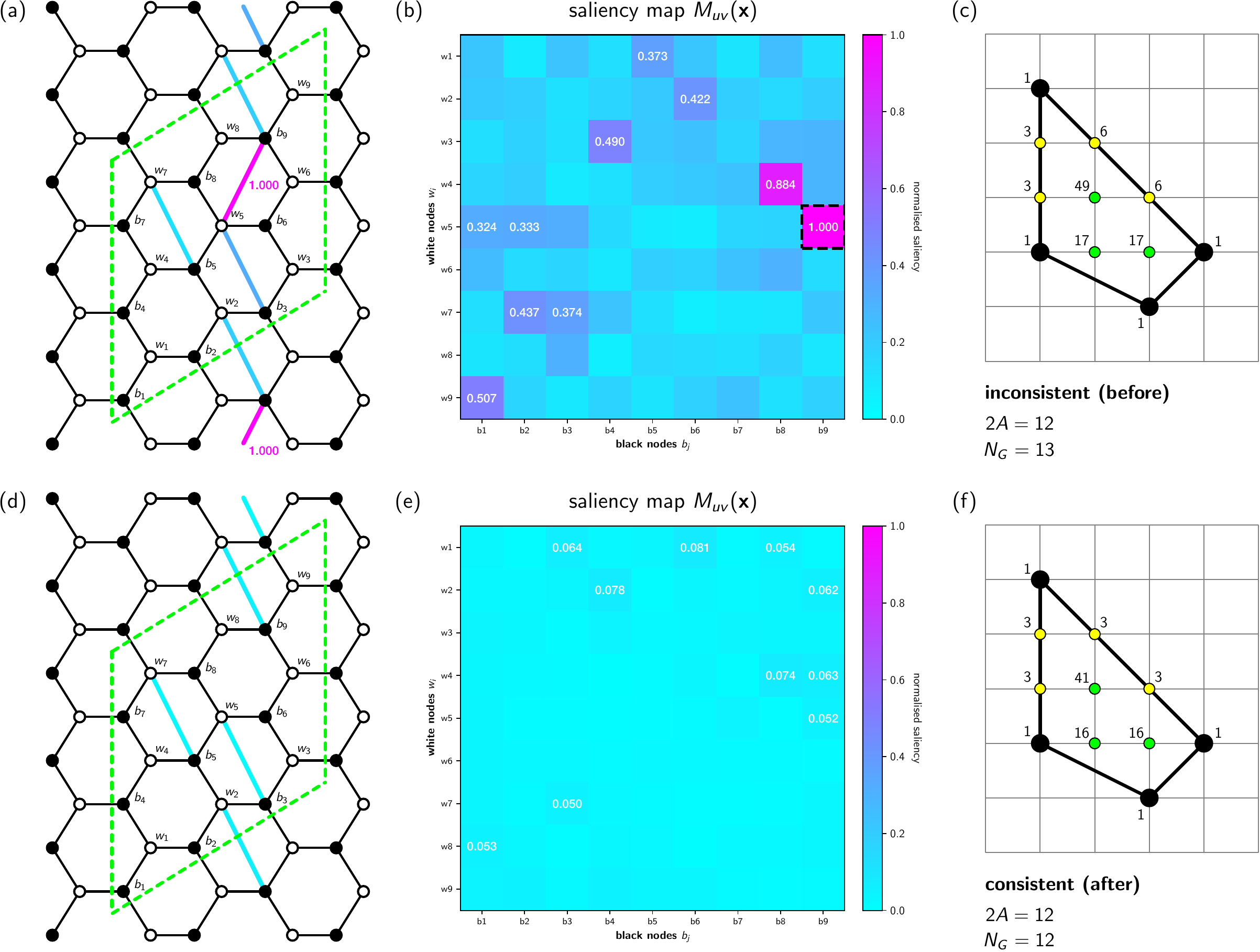} 
}
\caption{
Example of a saliency-guided diagnostic and repair of a representative $1$-edge-fixable brane tiling.
(a)--(c) The inconsistent brane tiling, its saliency map $M_{uv}(\mathbf{x})$, 
and the corresponding toric diagram, for which $N_G=13\neq2A=12$. 
The maximum saliency identifies the diagonal edge connecting $(w_5,b_9)$.
(d)--(f) Removing this edge via Higgsing the corresponding bifundamental chiral field in the $4d$ $\mathcal{N}=1$ theory 
yields a consistent brane tiling with $N_G=2A=12$.
\label{fig_07}}
 \end{center}
 \end{figure*}

\paragraph{Evaluation Set.}
We test whether the saliency tensor $\mathcal{S}(\mathbf{x})$
obtained from the trained CNN localizes the diagonal edges responsible
for inconsistency.
In our work, we focus only on the diagonal edges added to the original hexagonal brane tiling 
for $\mathbb{C}^3/\mathbb{Z}_3 \times \mathbb{Z}_3$ 
with orbifold action $(0,1,2)(1,0,2)$.
We call an added diagonal edge a \textit{defect} if removing that
edge alone restores consistency. A brane tiling containing at least one such edge is called \textit{$1$-edge-fixable}.
Among the $260{,}757$ inconsistent brane tilings in our dataset, we have $7{,}236$ brane tilings that are $1$-edge-fixable.
Amongst these, we have $5{,}157$ brane tilings that have a unique defect 
and $2{,}079$ that contain two defects.
In this dataset of $7{,}236$ brane tilings,
the number of added diagonal edges has a mean of $6.43$ as summarized in \fref{fig_05}.

\paragraph{Representative Examples.}
Let us consider the inconsistent brane tiling shown in \fref{fig_06}(a).
The brane tiling has $N_G=13$ while the corresponding toric diagram has $2A = 12$ as shown in \fref{fig_06}(c).
The largest component of the saliency tensor
$\mathcal{S}_{uvk}(\mathbf{x})$ corresponds to the diagonal edge connecting $(w_3,b_7)$.
Under the saliency map used for visualization, this component of the saliency tensor gives the
normalized saliency map value $M_{37}(\mathbf{x})=1.000$.
The diagonal edge connecting $(w_3,b_7)$ 
corresponds to a bifundamental chiral field in the associated $4d$ $\mathcal{N}=1$ theory.
Higgsing this field removes the edge and leads to a consistent brane tiling with
$N_G=12$ and $2A = N_G=12$, while leaving the shape and area $A$ of the associated toric diagram unchanged. 

\begin{figure*}[htt!!]
\begin{center}
\resizebox{0.95\hsize}{!}{
\includegraphics[height=5cm]{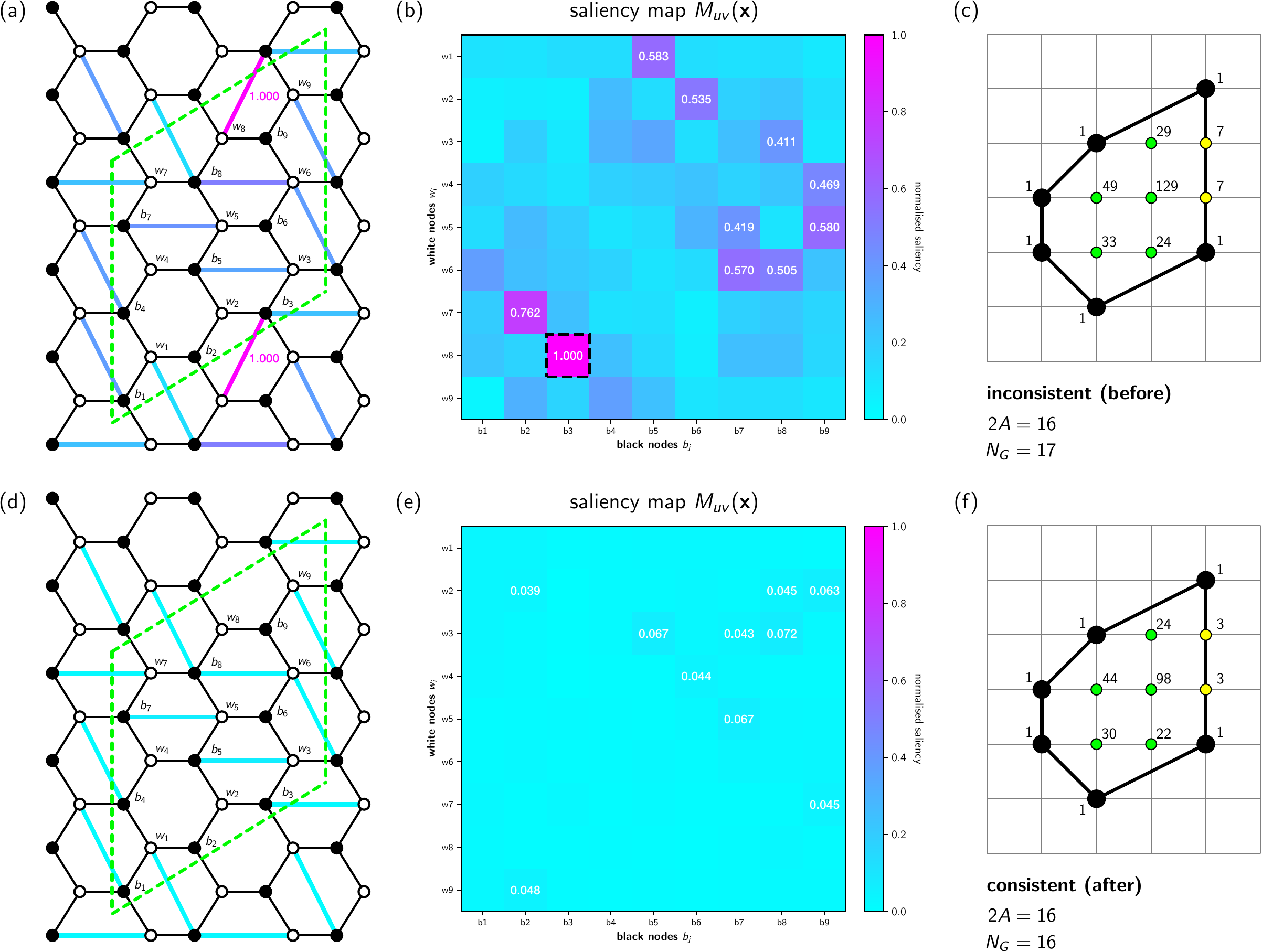} 
}
\caption{
Another example of a saliency-guided diagnostic and repair of a representative $1$-edge-fixable brane tiling.
(a)--(c) The inconsistent brane tiling, its saliency map $M_{uv}(\mathbf{x})$, 
and the corresponding toric diagram, for which $N_G=17\neq2A=16$. 
The maximum saliency identifies the diagonal edge connecting $(w_8,b_3)$.
(d)--(f) Removing this edge via Higgsing the corresponding bifundamental chiral field in the $4d$ $\mathcal{N}=1$ theory 
yields a consistent brane tiling with $N_G=2A=16$.
\label{fig_08}}
 \end{center}
 \end{figure*}

Two further examples are shown in 
\fref{fig_07} and \fref{fig_08}.
In both cases, 
removing the diagonal edge with the highest saliency restores consistency without changing the shape and area of the corresponding toric diagram.

\paragraph{Localization Accuracy.}
For each brane tiling, we rank the
$729$ components
of the saliency tensor $\mathcal{S}_{uvk}(\mathbf{x})$.
A Top-$k$ localization is successful if at least one of the $k$
highest ranked components corresponds to a defect whose removal restores
consistency of the brane tiling.
Thus, Top-$1$ requires the highest ranked saliency tensor component to identify a defect, 
whereas Top-$3$ requires a defect to occur among the $3$ highest ranked components of the saliency tensor.
Unless otherwise stated, the ranking is performed over all $729$
components, including those for which the corresponding input Kasteleyn tensor component $\mathbf{x}_{uvk}$ is zero.

Using an unrestricted ranking over all $729$ components of the saliency tensor 
defined in \eref{es05a10},
the trained CNN achieves a Top-$1$ and Top-$3$ 
defect localization accuracy of $82.92\%$ and $92.27\%$, respectively, 
over the $7{,}236$ $1$-edge-fixable brane tilings.
We note here that whenever the highest ranked component of the saliency tensor
corresponds to an edge present in the input brane tiling, that edge is always a diagonal edge.
None of the $7{,}236$ $1$-edge-fixable brane tilings has its maximum saliency on an edge
belonging to the original hexagonal brane tiling of $\mathbb{C}^3/\mathbb{Z}_3 \times \mathbb{Z}_3$ with orbifold action $(0,1,2)(1,0,2)$.

For $6{,}667$ of the $7{,}236$ $1$-edge-fixable brane tilings,
the highest ranked saliency tensor component corresponds to a diagonal edge present in the input brane tiling.
For the remaining $569$ brane tilings, its corresponding input tensor component
$\mathbf{x}_{uvk}$ is zero and represents no edge present in the input brane tiling.
When we restrict the ranking of saliency tensor components
to those corresponding to diagonal
edges that are actually present in the input brane tiling, 
then the Top-$1$ localization accuracy rises to $88.18\%$ and the Top-$3$ localization accuracy rises to $99.32\%$.

For the same restricted set of candidate edges,
a uniformly random choice among the $N_D$ diagonal edges
present in a brane tiling
has defect localization success probability $N_R/N_D$, 
where $N_R$ is the number of defects.
Averaging this probability over the evaluation set
gives a matched Top-$1$ random baseline of $21.10\%$.
The restricted Top-$1$ accuracy of $88.18\%$ is therefore more than $4$ times higher than the matched random baseline. 
We therefore conclude that the saliency tensor obtained from the trained CNN
contains substantial information about the location of defects in the input brane tiling.

\paragraph{Dependence on the Number of Diagonal Edges.}
\tref{tab_01} summarizes the defect localization accuracy obtained from
the saliency tensor according to the number of diagonal edges $N_D$ added to the $1$-edge-fixable brane tilings in the evaluation set. 
Here, we note that for $N_D=1$ there are $27$ brane tilings, which are all inconsistent and are $1$-edge-fixable.
Furthermore, for $N_D=2$ and $N_D=3$, there are no $1$-edge-fixable brane tilings. 

We observe that the localization performance of the saliency tensor is not monotonic in $N_D$.
Brane tilings with $N_D=4$ and $N_D=7$ reach Top-$1$ localization accuracies of $91.15\%$
and $92.11\%$, respectively, whereas the $N_D=6$ and $N_D=9$ brane tilings reach $59.57\%$ and $60.74\%$, respectively.
We note here that the difference in performance cannot be explained by the number of defects $N_R$.
For example, every brane tiling with $N_D=1$ and every brane tiling with $N_D=6$ has exactly one defect,
yet the Top-$1$ localization accuracies of these two sets are $96.30\%$ and $59.57\%$, respectively.
Conversely, the $N_D=7$ and $N_D=9$ brane tilings contain 
similar proportions of brane tilings with
$N_R=1$ and $N_R=2$, 
but their Top-$1$ localization accuracies are very different. 
Moreover, having $N_R=2$ defects should make localization of defects easier rather than harder
since a prediction is counted as correct if either defect receives the highest saliency. 

Instead, the variation of localization accuracy
broadly follows the mean defect fraction $\langle N_R / N_D \rangle$, 
which provides a measure of localization difficulty.
We observe that for $N_D=6$ and $N_D=9$, the mean defect fraction is $16.67\%$ and $15.56\%$, respectively.
Thus, defects constitute a relatively small fraction of the available diagonal edges for these cases,
which leaves fewer correct candidates for defect localization.
Because the Top-$1$ accuracies in \tref{tab_01} 
are obtained from the unrestricted ranking over all $729$ saliency tensor components, 
the mean defect fraction should be interpreted as a difficulty descriptor rather than as a directly matched random baseline.

For brane tilings with $N_D=6$, 
another contributor to the drop in performance is that
many of the highest saliency components in the input tensor $\mathbf{x}$ correspond to no edge present in the input brane tiling. 
This occurs in $15.12\%$ of the $N_D=6$ brane tilings compared to the $7.86\%$ over the full evaluation set. 
If the saliency ranking is restricted to diagonal edges that are actually present in the brane tiling, 
then the Top-$1$ localization accuracy for $N_D=6$ increases from $59.57\%$ to $64.51\%$.
Moreover, the Top-$3$ accuracy is considerably higher than the Top-$1$ accuracy, reaching $69.75\%$ for $N_D=6$ and $71.11\%$ for $N_D=9$.
Accordingly, we note here that even when the highest saliency is not assigned to the correct defect in the brane tiling, 
the defect is often 
among the candidate diagonal edges that are ranked highest by saliency.

\paragraph{Dataset Exposure.}
We test whether the defect localization performance of the trained CNN depends on whether 
a brane tiling appeared in the training or validation set.
Among the $7{,}236$ brane tilings in the evaluation set,
$4{,}623$ appeared in either the training set ($4{,}162$) or the validation set ($461$),
whereas the remaining $2{,}613$ appeared in neither set.
We refer to these two groups as the \textit{exposed} and \textit{unexposed} brane tilings, respectively.
The distributions of the number of diagonal edges $N_D$ are similar for the exposed and unexposed brane tiling sets.
Therefore, the comparison is not biased by either set containing systematically more complicated brane tilings for the localization task.
Moreover, no defect location labels were used during training or validation
and the CNN was provided only the binary consistency labels $y=0$ and $y=1$ for inconsistent and consistent brane tilings, respectively.

We observe that the localization accuracies for the two sets of brane tilings are very similar.
For the unexposed brane tilings, the Top-$1$ and Top-$3$ localization accuracies are $82.17\%$ and $91.85\%$, respectively, 
while for the exposed brane tilings, the Top-$1$ and Top-$3$ localization accuracies are $83.34\%$ and $92.52\%$.
Accordingly, we observe that prior exposure of the brane tilings only changes the Top-$1$ and Top-$3$ accuracies by $1.17$ and $0.67$ percentage points, respectively.
The CNN therefore achieves nearly the same defect localization performance on brane tilings that were not used for training and validation. 
Together with the fact that no defect locations were supplied for the CNN, 
the comparable localization performance on unexposed brane tilings indicates that 
successful localization of defects does not require memorization of individual brane tilings in the training or validation sets.

\section{Conclusions and Discussion}

In this work, we introduced a convolutional neural network (CNN) that distinguishes geometrically consistent from geometrically inconsistent brane tilings 
and their associated $4d$ $\mathcal{N}=1$ supersymmetric gauge theories.
To our knowledge, this is the first application of explainable AI
to the diagnosis and resolution of inconsistencies in quantum field theories arising in string theory. 
For the family of $262{,}144$ brane tilings studied here, 
the CNN learns geometric consistency directly from the Kasteleyn matrix of the brane tiling. 
Moreover, 
gradient-based saliency localizes the responsible defect chiral fields
with a Top-$3$ accuracy of up to $99.32\%$ when the ranking is restricted to the diagonal edges present in the brane tiling.

The broader significance is that the trained CNN achieves 
more than a classification of quantum field theories 
into consistent and inconsistent classes. 
Although geometric consistency is determined by global properties of the corresponding brane tiling, 
the trained neural network associates its prediction with particular local constituents of the brane tiling and of the associated $4d$ $\mathcal{N}=1$ theories.
This localization performance persists for brane tilings and the corresponding Kasteleyn tensors that were never encountered during the training of the CNN. 
The neural network therefore learns information about the structure underlying geometric consistency that can be used 
both to diagnose an inconsistent $4d$ $\mathcal{N}=1$ theory and to suggest remedies to restore its consistency. 

It would be interesting to determine how far this explainable machine learning framework 
extends beyond the family of $4d$ $\mathcal{N}=1$ supersymmetric gauge theories realized by brane tilings \cite{Franco:2015tna, Franco:2015tya, Franco:2016nwv, Franco:2016qxh, Franco:2016fxm, Franco:2016tcm, Franco:2022gvl, Franco:2023flw, Ghim:2024asj, Kho:2025fmp, Kho:2025jxk, Kho:2026zwc, Kwon:2026stu}.
We also note that more refined explainable AI methods may improve the localization of defects in large families of supersymmetric gauge theories realized in string theory.
We plan to report on these studies in future work.

\subsection*{Acknowledgements}

S.-J. L. is supported by the Institute for Basic Science (IBS) under Project No. IBS-R003-D1.
R.-K. S. is supported by an Outstanding Young Scientist Grant (RS-2025-00516583) of the National Research
Foundation of Korea (NRF). He is also partly supported by the BK21 Program (“Next Generation Education Program for Mathematical Sciences”, 4299990414089) 
funded by the Ministry of Education in Korea and the National Research Foundation of Korea (NRF).
The authors would like to thank 
the Shanghai Institute for Mathematics and Interdisciplinary Sciences (SIMIS), 
the Simons Center for Geometry and Physics at Stony Brook University, 
the Korea Institute for Advanced Study (KIAS),
as well as the Beijing Institute of Mathematical Sciences and Applications (BIMSA),
where the final stages of this work were conducted. 

\bibliographystyle{jhep}
\bibliography{mybib}

\end{document}